\documentclass[twocolumn,showpacs,preprintnumbers,aps]{revtex4}
\usepackage{amssymb}
\usepackage{amsmath}
\usepackage{graphicx}

\input{tcilatex}

\begin{document}

\title{Instability of coherent states \ of a real scalar field}
\author{Vladimir A. Koutvitsky}
\affiliation{Pushkov Institute of Terrestrial Magnetism, Ionosphere and Radiowave
Propagation of the Russian Academy of Sciences (IZMIRAN), Troitsk, Moscow
Region, 142190, Russia}
\author{Eugene M. Maslov}
\affiliation{Pushkov Institute of Terrestrial Magnetism, Ionosphere and Radiowave
Propagation of the Russian Academy of Sciences (IZMIRAN), Troitsk, Moscow
Region, 142190, Russia}

\begin{abstract}
We investigate stability of both localized time-periodic coherent states
(pulsons) and uniformly distributed coherent states (oscillating condensate)
of a real scalar field satisfying the Klein-Gordon equation with a
logarithmic nonlinearity. The linear analysis of time-dependent parts of
perturbations leads to the Hill equation with a singular coefficient. To
evaluate the characteristic exponent we extend the Lindemann-Stieltjes
method, usually applied to the Mathieu and Lam\'{e} equations, to the case
that the periodic coefficient in the general Hill equation is an unbounded
function of time. As a result, we derive the formula for the characteristic
exponent and calculate the stability-instability chart. Then we analyze the
spatial structure of the perturbations. Using these results we show that the
pulsons of any amplitudes, remaining well-localized objects, lose their
coherence with time. This means that, strictly speaking, all pulsons of the
model considered are unstable. Nevertheless, for the nodeless pulsons the
rate of the coherence breaking in narrow ranges of amplitudes is found to be
very small, so that such pulsons can be long-lived. Further, we use the
obtaned stability-instability chart to examine the Affleck-Dine type
condensate. We conclude the oscillating condensate can decay into an
ensemble of the nodeless pulsons.
\end{abstract}

\maketitle

\section{INTRODUCTION}

Nonlinear localized field configurations, solitons, are currently considered
as models of various physical objects, from elementary particles and collective
excitations in condensed matter to giant
lumps of dark matter in the form of soliton stars and galactic halos \cite%
{Bishop, Lee, Mielke}. Stability properties of solitons were investigated by many authors,
and a number of important results has been obtained (see, e.g., \cite%
{Rybakov} and references therein). In particular, Hobart \cite{Hobart} and
Derrick \cite{Derrick} have proved that static multidimensional scalar
solitons are energetically unstable, and, hence, these objects cannot last
in a real world for a long time. One way to avoid this theorem is to invoke
time dependence. Along this line main efforts were focused on the stability
analysis of the stationary states, i.e., coherent states of a complex scalar
field oscillating harmonically in time. It turned out, however, that for a
wide class of relativistic models these states can be only conditionally
stable, i.e., stable with respect to a certain type of perturbations (e.g.,
conserving the scalar charge) \cite{Rybakov}. As to the time-periodic states
of a more general form, both complex and real, there are presently no strong
analytical results on their stability.

In this paper we examine stability of time-periodic configurations of the
form%
\begin{equation}
\phi =\phi _{0}(t,\mathbf{r})=a(t)u(\mathbf{r})  \label{eq2}
\end{equation}%
satisfying the nonlinear Klein-Gordon equation 
\begin{equation}
\phi _{tt}-\Delta \phi +U^{\prime }(\phi )=0.  \label{eq1}
\end{equation}%
These are coherent states in the sense the field oscillates synchronously at
all spatial points. It is necessary to stress that we consider real
solutions, so that the energy density oscillates as well (in contrast to the
stationary states for which $a(t)$\ $\varpropto e^{i\omega t}$). Solitons
with oscillating energy density are usually called pulsons.

It turns out that for real $\phi $ the ansatz (\ref{eq2}) determines
uniquely the potential $U(\phi )$ in Eq. (\ref{eq1}). Namely, if neither $%
a(t)$ nor $u(\mathbf{r})$ are constants, the only potential admitting such
solutions will have the form \cite{Maslov1}%
\begin{equation}
U(\phi )=\frac{1}{2}\phi ^{2}[m^{2}+\lambda (1-\ln \phi ^{2})],  \label{eq3}
\end{equation}%
where $m^{2}$ and $\lambda $ are arbitrary constants.

Originally, the Klein-Gordon equation with a logarithmic potential of this
type has been introduced in the quantum field theory by G. Rosen \cite{Rosen}%
. Later on Bialynicki-Birula and Mycielski \cite{B-BM1} have rediscovered
this equation and also considered its nonrelativistic version, the nonlinear
Schr\"{o}dinger equation \cite{B-BM2}.

In inflationary cosmology and in modern supersymmetric field theories the
logarithmic nonlinearities appear naturally when quantum corrections to
effective potentials are allowed for \cite{Linde1, Linde2, Barrow, Enqvist1}%
. In this context the expression in the square brackets of Eq. (\ref{eq3})
can be treated as a dynamic inflaton mass term $m_{S}^{2}$ that is the bare
inflaton mass term $m^{2}$ plus the logarithmic correction. It can be
represented in the commonly considered form \cite{Enqvist1, Enqvist4,
Multamaki1, Multamaki2, Kasuya, Enqvist2, Enqvist3, Pawl} by the substitution $\ln
\phi ^{2}=1+\ln (\Phi /M)^{2}$, $\lambda /m^{2}=-K$, where $\Phi $ is an
inflaton scalar field, $M$ is a large mass scale, $K$ is a constant (usually
negative and small). Thus our consideration is also relevant to dynamics of
the pulson exitations of a real inflaton field oscillating around a vacuum
value.

Note that the multidimensional pulsons probably exist in other scalar models
as well. Thus the long-living oscillating spherically symmetric localized
states were numerically found in the sine-Gordon, $\phi ^{4}$, and $\phi
^{3}-\phi ^{4}$ models \cite{Bog-Mak1,Bog,Bog-Mak2,Gleiser1,Copeland} (see %
\cite{Belova}\ for a review). Unfortunately, the analytic form of these
solutions is so far unknown.

The model (\ref{eq1}) and (\ref{eq3}) is unique in the sense it has a whole
family of exact pulson solutions of the form (\ref{eq2}), all existing in
any number of spatial dimensions \cite{Maslov1}. This is also true for
complex version of the model \cite{Marques,Bogolubsky}. The real pulsons we
are dealing with are the limiting states of the complex ones, when the
scalar charge tends to zero. Other limiting states are Q-balls for which $%
a(t)\varpropto e^{i\omega t}$ \cite{Enqvist4, Multamaki1}. It is believed
that Q-balls can arise due to fragmentation of the Affleck-Dine condensate %
\cite{Kasuya, Enqvist2, Enqvist3, Pawl}. We will see below that the parametric
instability of the oscillating condensate leads to the resonant
fragmentation that can give rise to the pulson formation at the nonlinear
stage. Like Q-balls \cite{Multamaki2}, pulsons interact elastically or
inelastically in collisions depending on their relative velocities, phases,
and rest masses \cite{Maslov2,Maslov3}. Thus, in model (\ref{eq1}) and (\ref%
{eq3}) the light pulsons with given relative velocities interact always
elastically, independently of their phases. In contrast, the collisions of
heavy pulsons can result in formation of the so-called explosons, localized
states with exponentially growing amplitude \cite{Maslov2}. For the
intermediate masses the picture depends essentially on the phases of the
colliding pulsons and impact velocity determining the duration of the
interaction \cite{Maslov3}.

The above results suggest that there is a domain of parameters where pulsons
are stable, at least in short time interactions. But in what sense? How long
a pulson conserves its characteristic features once interaction ends? If
pulsons are long-lived objects they will be interesting candidates for the
dark matter constituents having time-dependent density. What is known about
stability of an isolated pulson at the long time scale? Surprisingly, but
very few. In Ref. \cite{Bogolubsky} it was argued in favour of its perfect
stability. No deviations from the exact solution (\ref{eq2}) were found
after about one thousand oscillations. However, our preliminary numerical
experiments \cite{Koutvitsky, Koutvitsky-Maslov} have shown that the pulsons
of certain amplitudes, even perturbed by computer round-off errors only,
gradually lose their coherency, remaining well-localized oscillating
objects. This has motivated the closer examination.

In the present paper we clarify how long the pulsons can conserve the
coherency depending on their parameters. For this purpose we investigate
stability of the spherically symmetric pulson solutions (\ref{eq2}) with
respect to small initial perturbations of an arbitrary form.

The paper is organized as follows. In Sec. II the main properties of the
real pulsons of the model considered are reviewed. Section III is wholly
devoted to the linear stability analysis. We arrive at the singular Hill
equation and generalize the Lindemann-Stieltjes method to evaluate the
characteristic exponent. On this basis we examine stability of the pulsons
and discuss fragmentation of the oscillating Affleck-Dine type condensate.
In Sec. IV we make some remarks concerning the complex pulsons and summarize
the main results.

\section{PULSONS AS COHERENT STATES}

Assuming $\lambda $ positive, let us first eliminate the constants $m^{2}$
and $\lambda $ from consideration by the scaling $t\rightarrow \lambda
^{-1/2}t$, $\mathbf{r}\rightarrow \lambda ^{-1/2}\mathbf{r}$, $\phi
\rightarrow \phi \exp \frac{m^{2}}{2\lambda }$. In the new variables the
field $\phi $ may be thought of as satisfying Eq. (\ref{eq1}) with the
potential

\begin{equation}
U(\phi )=\frac{1}{2}\phi ^{2}(1-\ln \phi ^{2}).  \label{eq4}
\end{equation}%
It is the potential we will deal with. It has local minimum at $\phi =0$ and
two maxima at $\phi =\pm 1$, at the minimum the potential having the
singularity: its second derivative tends to infinity as $\phi \rightarrow 0$.

The substitution of the ansatz (\ref{eq2}) into Eq. (\ref{eq1}) leads then
to two independent equations, 
\begin{equation}
a_{tt}=-\frac{d}{da}\left[ \frac{1}{2}a^{2}(1-\ln a^{2})\right] ,
\label{eq5}
\end{equation}%
\begin{equation}
\Delta u=-\frac{d}{du}\left[ \frac{1}{2}u^{2}(\ln u^{2}-1)\right] .
\label{eq6}
\end{equation}%
Note that the potentials in the square brackets of Eqs. (\ref{eq5}) and (\ref%
{eq6}) have the same form as the potential (\ref{eq4}) taken with plus and
minus signs, respectively. The existence of the oscillating localized
solutions (\ref{eq2}) is thus apparent from consideration of motion of a
mechanical particle in these potentials.

Let us consider in more detail the oscillatory solutions of Eq. (\ref{eq5}).
Using the Hamiltonian and denoting $\xi =a/a_{\max }$ ($0<a_{\max }<1$, $%
-1\leqslant \xi \leqslant 1$), we obtain%
\begin{equation}
\xi _{t}^{2}=\omega _{0}^{2}(1-\xi ^{2})+\xi ^{2}\ln \xi ^{2},  \label{eq7}
\end{equation}%
where 
\begin{equation}
\omega _{0}^{2}=1-\ln a_{\max }^{2}>1.  \label{eq8}
\end{equation}%
In the case of small amplitudes, $a_{\max }^{2}\ll 1$, $\omega _{0}^{2}\gg 1$%
, Eq. (\ref{eq7}) gives 
\begin{equation}
\xi (t)\approx \cos \omega _{0}t.  \label{eq9}
\end{equation}%
Thus, we have quasi-harmonic high-frequency oscillations which are however
nonlinear since their period, 
\begin{equation}
T\approx \frac{2\pi }{\left| \ln a_{\max }^{2}\right| ^{1/2}},  \label{eq10}
\end{equation}%
depends on the amplitude \cite{Koutvitsky-Maslov}. In the next approximation
from Eq. (\ref{eq7}) we find%
\begin{equation}
T=\frac{2\pi }{\omega _{0}}\left( 1+\frac{0.307}{\omega _{0}^{2}}+O\left( 
\frac{1}{\omega _{0}^{4}}\right) \right) .  \label{eq11}
\end{equation}%
In the case of near-critical amplitudes, when $a_{\max }^{2}\rightarrow 1$,
the oscillations become almost rectangular and have the period%
\begin{equation}
T\approx 2\sqrt{2}\ln \frac{1}{1-a_{\max }^{2}}.  \label{eq12}
\end{equation}%
Examples of solutions of Eq. (\ref{eq5}) are shown in Fig. 1. 
\begin{figure}[tbp]
\includegraphics[width=8cm]{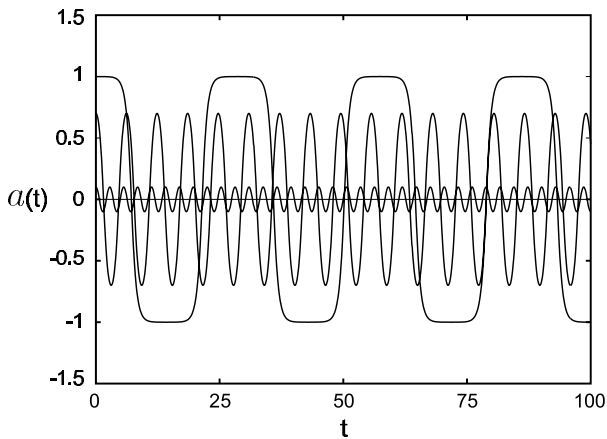}
\caption{Oscillatory solutions of Eq. (\ref{eq5}) for $a_{\max }=0.1$, $%
a_{\max }=0.7$, and $a_{\max }=0.9999$.}
\end{figure}

The spatial structure of a pulson is determined by Eq. (\ref{eq6}). In the
spherically symmetric case this equation has a discrete spectrum of
localized $N$-nodal solutions $u_{N}(r)$ with the first derivatives
vanishing at the origin \cite{B-BM3} (see Fig. 2). The simplest of them, the
nodeless solution, has a Gaussian-like shape, 
\begin{equation}
u_{0}(r)=e^{(3-r^{2})/2},  \label{eq13}
\end{equation}%
and is usually called gausson \cite{Rosen,B-BM1,B-BM2}. It is agreed that
its effective radius equals $\sqrt{2}$. In the multinodal solutions, as $r$
increases, the field undergoes spatial oscillations of the half-wavelength $%
L\lesssim 2\sqrt{2}$ and then decays as%
\begin{equation}
u_{N}(r)\approx C_{N}e^{-(r-\rho _{N})^{2}/2}\quad (r\gg r-\rho _{N},\;N\gg
1),  \label{eq13a}
\end{equation}%
where $C_{N}$ is the value of the last extremum of $u_{N}(r)$ attained at $%
r=\rho _{N}$, $C_{N}\rightarrow (-1)^{N}e^{1/2}\;(N\rightarrow \infty )$, $%
\rho _{N}\sim NL$. Thus, the pulsons of the model (\ref{eq1}) and (\ref{eq4}%
) are well-localized states of the inhomogeneity length $L$, at all points
the field oscillating coherently with the period $T$. (To return to the
physical units these scales should be multiplied by $\lambda ^{-1/2}$.) In
our dimensionless variables the pulsons are characterized by two parameters
only: the amplitude $a_{\max }$ and the number of the nodes $N$ (or $T$ and $%
u_{N}(0)$, respectively). 
\begin{figure}[th]
\includegraphics[width=8cm]{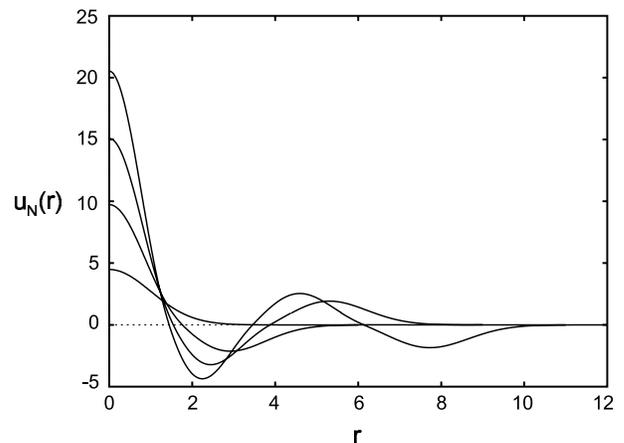}
\caption{Spectrum of the spherically symmetric $N$-nodal solutions $u_{N}(r)$
of Eq. (\ref{eq6}): $u_{0}(0)=e^{3/2}$, $u_{1}(0)=9.726$, $u_{2}(0)=15.084$, 
$u_{3}(0)=20.526$, \dots\ .}
\end{figure}

It should be stressed that, due to nonanalyticity of $U(\phi )$ at $\phi =0$%
, the right-hand sides of Eqs. (\ref{eq5}) and (\ref{eq6}) are nonanalytic
when $a$ and $u$ become zero. Hence, the solutions $a(t)$ and $u(r)$
themselves become nonanalytic at those points $t$ and $r$ where they pass
through zero. Thus in the solution (\ref{eq9}) we have dropped the terms
which are small (of the order of $\omega _{0}^{-2}$) but nonanalytic when $%
\xi (t)=0$. In general case from Eq. (\ref{eq7}) it follows that $\xi (t)$
passes through zero at $t=t_{m}$ as%
\begin{eqnarray}
\underset{t\rightarrow t_{m}}{\xi (t)} &=&\pm \omega _{0}(t-t_{m})\Bigl[1+%
\frac{1}{6}(t-t_{m})^{2}\ln (t-t_{m})^{2}  \notag \\
&&+O\left( (t-t_{m})^{2}\right) \Bigr],  \label{eq14}
\end{eqnarray}
where $\pm $ sign is taken for $\xi _{t}(t_{m})\gtrless 0$. It is seen that $%
\xi _{ttt}$ becomes infinite as $t\rightarrow t_{m}$. Similarly, one can
show that in the vicinity of the $n$-th node ($n=1,\dots,N$)%
\begin{eqnarray}
\underset{r\rightarrow r_{n}}{u(r)} &=&u_{r}(r_{n})(r-r_{n})\Bigl[1-\frac{%
r-r_{n}}{r_{n}}  \notag \\
&&-\frac{1}{6}{(r-r_{n})^{2}\ln (r-r_{n})^{2}}  \notag \\
&&+O\left( (r-r_{n})^{2}\right)\Bigr].  \label{eq15}
\end{eqnarray}
As we will see below, the nonanalyticity of $U(\phi )$ gives rise to some
specific features of the stability analysis.

\section{THE LINEAR STABILITY ANALYSIS}

Consider a small fluctuation $\eta (t,\mathbf{r})$ around the spherically
symmetric pulson (\ref{eq2}), $\phi =\phi _{0}(t,r)+\eta (t,\mathbf{r})$. In
the linear approximation the equation for $\eta $ reads%
\begin{equation}
\eta _{tt}-\Delta \eta -(2+\ln \phi _{0}^{2})\eta =0.  \label{eq17}
\end{equation}%
Seeking a solution in the form $\eta (t,\mathbf{r})\varpropto X(t)\Psi (%
\mathbf{r})$ we arrive at the equations%
\begin{eqnarray}
X_{tt}+(E-2-\ln a^{2})X &=&0,  \label{eq19} \\
\Delta \Psi +(E+\ln u^{2})\Psi &=&0,  \label{eq20}
\end{eqnarray}%
where $E$ is some constant.

The expression in the brackets of Eq. (\ref{eq17}) is $-U^{\prime \prime
}(\phi _{0})$. It becomes infinite, as well as the expressions in the
brackets of Eqs. (\ref{eq19}) and (\ref{eq20}), at the points $t_{m}$ and $%
r_{n}$ where $a(t)$ and $u(r)$ become zero. Thus we need to analyze the
second order differential equations with singular coefficients. We begin
with Eq. (\ref{eq19}) which has the periodic singular coefficient $\ln
a^{2}(t)$ and hence belongs to the class of Hill equations.

\subsection{Singular Hill equation and generalized Lindemann-Stieltjes method%
}

It turns out to be very useful to look at the problem as a whole,
considering first the Hill equation of a general form 
\begin{equation}
X_{tt}+h(z(t))X=0.  \label{eq21}
\end{equation}%
We will assume that $h(z)$ is an integral function of $z$, while $z(t)$ is a
real-valued periodic (of a period $\tau $) even function of $t$, having, in
general, singularities, but such, that $h(z(t))$ remains still integrable.

It is well known that the Hill equation describes the physical systems in
which the parametric resonance can occur. In the context of our stability
analysis we will be interested in real resonant solutions of Eq. (\ref{eq21}%
). In accordance with the Floquet theory (see, e.g., \cite{Magnus}), any one
of these solutions can be represented as a linear combination of the
fundamental solutions 
\begin{equation}
X_{+}(t)=\varphi (t)e^{\mu t},\quad X_{-}(t)=\varphi (-t)e^{-\mu t},
\label{eq22}
\end{equation}%
where $\varphi (t)$ is a $\tau $-periodic or $\tau $-antiperiodic real
function, $\mu >0$ is the characteristic exponent. In the case that $z(t)$
is unbounded it is impossible to obtain the solutions and evaluate $\mu $ by
expansions in Fourier series, following the standard Hill approach. Another
way is to apply the Lindemann-Stieltjes method \cite{Whittaker}. In some
cases it allows one to obtain the results in a closed analytical form \cite%
{Greene,Kaiser,Maslov4,Finkel}. We first outline this method in the context
of the general Hill equation (\ref{eq21}) with an extension to the case that
the periodic function $z(t)$ is unbounded. In doing so we follow the paper %
\cite{Maslov4} where the method was used to construct the resonant solutions
of the Lam\'{e} equation.

The main idea is as follows. Let us treate $z$ as a new \textquotedblleft
time\textquotedblright\ variable instead of $t$. In each interval of
monotonicity of $z(t)$ we define 
\begin{equation}
y(z)=X(t).  \label{eq24}
\end{equation}%
Assume that the periodic function $z(t)$ satisfies the equation%
\begin{equation}
z_{t}^{2}=g(z),  \label{eq23}
\end{equation}%
where $g(z)$ is an integral function of $z$. Eq. (\ref{eq21}) then becomes 
\begin{equation}
g(z)y^{\prime \prime }+\frac{1}{2}g^{\prime }(z)y^{\prime }+h(z)y=0
\label{eq25}
\end{equation}
(hereinafter the prime denotes $d/dz$).

Let us first suppose $z(t)$ is bounded. Equation (\ref{eq23}) then shows
that it is differentiable. Zeros of the function $g(z)$ on the complex $z$
plane, taken to be isolated, are singular points of Eq. (\ref{eq25}). Since $%
z(t)$ is periodic and real-valued, among singular points there are two, $%
\zeta _{1}$ and $\zeta _{2}$, lying on the real axis and being minimal and
maximal values which $z(t)$ acquires at the end points of the intervals of
monotonicity. Also, it follows that $g^{\prime }(\zeta _{1,2})\neq 0$.
Physically, this is well understood, since $\zeta _{1}$ and $\zeta _{2}$ can
be treated as turning points in periodic motion of a mechanical particle,
e.g., of a nonlinear oscillator, under the action of the force $g^{\prime
}/2 $. From Eq. (\ref{eq23}) it is clear that the interval $\left[ \zeta
_{1},\zeta _{2}\right] $ does not contain other singular points of Eq. (\ref%
{eq25}).

For example, in the case of the Mathieu equation we have $z(t)=\cos ^{2}t$, $%
g(z)=4z(1-z)$, so that Eq. (\ref{eq25}) has the regular singular points $%
z=\zeta _{1}=0$, $z=\zeta _{2}=1$, both being the turning points. In
addition, the equation has an irregular singularity at infinity. For the Lam%
\'{e} equation $z(t)=\func{sn}^{2}(t,\varkappa )$, $g(z)=4z(1-z)(1-\varkappa
^{2}z)$. Equation (\ref{eq25}) then has the regular singular points $z=\zeta
_{1}=0$, $z=\zeta _{2}=1$, $z=\varkappa ^{-2}>1$, first two of them being
the turning points, and a regular singularity at infinity.

In general, it is easy to verify that the turning points $\zeta _{1,2}$ are
regular singular points of Eq. (\ref{eq25}), the exponents at each being $0$
and $1/2$. This implies that in the vicinity of each turning point $\zeta $
there exist two independent solutions of Eq. (\ref{eq25}), $y^{(0)}(z;\zeta
) $ and $y^{(1/2)}(z;\zeta )$, having asymptotics $1+O(z-\zeta )$ and $%
(z-\zeta )^{1/2}[1+O(z-\zeta )]$, correspondingly.

Now let us consider any one interval of monotonicity of the $\tau $-periodic
even function $z(t)$. Denote as $y_{1}(z)$ and $y_{2}(z)$ those two linearly
independent solutions of Eq. (\ref{eq25}) one of which coinsides, by Eq. (%
\ref{eq24}), with the increasing solution (\ref{eq22}), $X_{+}(t)$, and
another with the decreasing one, $X_{-}(t)$, on the interval chosen. Since $%
\varphi (t)$ is either $\tau $-periodic or $\tau $-antiperiodic, the product 
$\varphi (t)\varphi (-t)=X_{+}X_{-}=y_{1}y_{2}=w(z)$ is always $\tau $%
-periodic even function defined on the whole $t$ axis. Hence, at the end
points of the intervals of monotonicity of $z(t)$, i.e., at $t_{m}=m\tau
/2\;(m=0,\pm 1,\dots)$, the derivative $[(X_{+}X_{-})_{t}]_{t=t_{m}}=0$ or,
what is the same, 
\begin{equation}
\left( w^{\prime }\sqrt{g}\right) _{z=\zeta _{1,2}}=0.  \label{eq26}
\end{equation}%
In the vicinity of a turning point $\zeta $ the solutions $y_{1}$ and $y_{2}$
can be represented as linear combinations of the solutions $y^{(0)}$ and $%
y^{(1/2)}$. Consequently, the singularity $(z-\zeta )^{1/2}$ is the only one
which the function $w(z)=y_{1}y_{2}$ might have. But its existence is in
contradiction with Eq. (\ref{eq26}), because $g^{\prime }(\zeta )\neq 0$
and, hence, $g(z)_{z\rightarrow \zeta }\sim g^{\prime }(\zeta )(z-\zeta )$.
Therefore, the product $y_{1}y_{2}$ is analytic at $z=\zeta _{1,2}$. Recall
now that the interval $\left[ \zeta _{1},\zeta _{2}\right] $ does not
contain other singular points of Eq. (\ref{eq25}) and the singular points
are assumed to be isolated. We thus conclude that on the complex $z$ plane
there exists a vicinity of the interval $\left[ \zeta _{1},\zeta _{2}\right] 
$, i.e., an open domain $D\supset \left[ \zeta _{1},\zeta _{2}\right] $, in
which $y_{1}y_{2}$ is an analytic function of $z$. In addition, it follows
that $y_{1}^{2}$ and $y_{2}^{2}$ of necessity have singularities of the type 
$(z-\zeta )^{1/2}$ and, thus, cannot satisfy Eq. (\ref{eq26}).

Now consider the case that one of the turning points or the both are at
infinity. This implies that at the corresponding instants $t_{m}$ the
functions $z(t)$ and $z_{t}(t)$ become unbounded, the latter changing the
sign. Nevertheless, we assume that in the vicinities of $t_{m}$ the function 
$h(z(t))$ in Eq. (\ref{eq21}) is integrable and $X$ is continuous, whence it
follows that $X_{t}$ and, therefore, $w_{t}$ are also continuous. Hence, as
before, $(w_{t})_{t=t_{m}}=0$ due to evenness and periodicity, so that we
arrive at Eq. (\ref{eq26}) again, where $\zeta _{1}=-\infty $ and/or $\zeta
_{2}=+\infty $.

It is easy to verify that the bilinear combinations $y_{1}^{2}$, $y_{1}y_{2}$%
, and $y_{2}^{2}$ constitute the fundamental system of solutions of the
third-order differential equation 
\begin{equation}
g(z)w^{\prime \prime \prime }+\frac{3}{2}g^{\prime }(z)w^{\prime \prime
}+\left( \frac{1}{2}g^{\prime \prime }(z)+4h(z)\right) w^{\prime
}+2h^{\prime }(z)w=0.  \label{eq27}
\end{equation}%
Equation (\ref{eq26}) is thus a common criterion for selection of the
solution%
\begin{equation}
w=y_{1}y_{2}  \label{eq28}
\end{equation}%
from the set of solutions of Eq. (\ref{eq27}). In the case that $z(t)$ is
bounded, Eq. (\ref{eq26}) is the equivalent to the requirement that a
solution of Eq. (\ref{eq27}) be analytic in $D$. If $z(t)$ is unbounded, Eq.
(\ref{eq26}) will give the boundary conditions at infinity which must be
satisfied in solving Eq. (\ref{eq27}). In this case $w(z)$ will be analytic
in a vicinity $D$ of one of the intervals $(-\infty ,\zeta _{2}]$, $[\zeta
_{1},\infty )$, $(-\infty ,\infty )$.

Thus, in a neighbourhood of any one point $\zeta \in D$ we can write the
expansions%
\begin{equation}
\begin{pmatrix}
w(z) \\ 
g(z) \\ 
h(z)
\end{pmatrix}%
=\sum\limits_{n=0}^{\infty }%
\begin{pmatrix}
w_{n} \\ 
g_{n} \\ 
h_{n}
\end{pmatrix}%
(z-\zeta )^{n},  \label{eq28a}
\end{equation}%
Substitution of (\ref{eq28a}) into Eq. (\ref{eq27}) leads to the following
set of equations for the coefficients:%
\begin{equation}
m\sum_{n=1}^{m+2}n(m+n)g_{m-n+2}w_{n}+4\sum_{n=0}^{m}(m+n)h_{m-n}w_{n}=0
\label{eq34}
\end{equation}%
$(m=1,2,\dots)$. Thus for $m=1$ we have%
\begin{equation}
6g_{0}w_{3}+3g_{1}w_{2}+(g_{2}+4h_{0})w_{1}+2h_{1}w_{0}=0.  \label{eq33}
\end{equation}%
Assuming $w(\zeta )\neq 0$, we normalize $w(z)$ by $w_{0}=1$. Then, at given 
$w_{1}$ and $w_{2}$ the remaining coefficients $w_{n}$ are determined from
Eqs. (\ref{eq34}). The choice of $w_{1}$ and $w_{2}$ is not arbitrary but
determined by Eq. (\ref{eq26}). Thus, setting $\zeta =\zeta _{1}$ and,
hence, $g_{0}=0$, we must choose $w_{1}$ in such a way that the series (\ref%
{eq28a}) for $w$ (or its continuation) converges at the second turning point 
$\zeta _{2}$, or satisfies the boundary condition at infinity (\ref{eq26})
if $\zeta _{2}=+\infty $. For the Mathieu and Lam\'{e} equations this leads
to the function $w(z)$ which is an integral one, for the latter it being a
polinomial \cite{Whittaker}. In these cases the domain $D$ is evidently the $%
z$ plane with $\left| z\right| <\infty $.

Let us suppose the function $w(z)$ (\ref{eq28}) is found. Return now to Eqs.
(\ref{eq21})-(\ref{eq25}). Denote as $W$ the Wronskian of the solutions (\ref%
{eq22}),%
\begin{equation}
X_{+}X_{-t}-X_{+t}X_{-}=W=\limfunc{const}.  \label{eq35}
\end{equation}%
Setting 
\begin{eqnarray}
y_{1} &=&X_{+},\quad y_{2}=X_{-}\qquad (z_{t}\geqslant 0),  \notag \\
y_{1} &=&X_{-},\quad y_{2}=X_{+}\qquad (z_{t}\leqslant 0),  \label{Eq36}
\end{eqnarray}%
we then obtain%
\begin{equation}
y_{1}y_{2}^{\prime }-y_{1}^{\prime }y_{2}=W/\sqrt{g},  \label{eq37}
\end{equation}%
where $\sqrt{g}\geqslant 0$ is assumed. The system of equations (\ref{eq28})
and (\ref{eq37}) can be easily solved, which gives%
\begin{equation}
y_{1,2}^{2}=\exp \int \frac{f_{\mp }}{w\sqrt{g}}\, dz,  \label{eq38}
\end{equation}%
where%
\begin{equation}
f_{\pm }=w^{\prime }\sqrt{g}\pm W.  \label{eq39}
\end{equation}

Now let us insert $y_{1,2}$ (\ref{eq38}) back into Eq. (\ref{eq25}). We
obtain%
\begin{equation}
2gww^{\prime \prime }+g^{\prime }ww^{\prime }-gw^{\prime 2}+4hw^{2}+W^{2}=0.
\label{eq40}
\end{equation}%
By this formula one can find the constant $W^{2}$ from a knowledge of $w(z)$
in a vicinity of any point $z$. Thus, calculating (\ref{eq40}) at any one
finite turning point $\zeta $ we obtain (in terms of expansions (\ref{eq28a}%
) with normalization $w_{0}=1$)%
\begin{equation}
W^{2}=-4h_{0}-g_{1}w_{1}.  \label{eq41}
\end{equation}%
Alternatively, one can take zeros $z_{i}$ of $w(z)$. (The functions $g(z)$
and $w(z)$ do not have common zeros because otherwise the Wronskian (\ref%
{eq35}) would be zero.) Then we find 
\begin{equation}
W^{2}=g(z_{i})w^{\prime 2}(z_{i}).  \label{eq42}
\end{equation}%
The requirement for positivity of $W^{2}$ determines the values of
parameters of Eq. (\ref{eq21}) (resonance zones) for which the resonant
solutions exist.

Let us construct these solutions. Consider the intervals of monotonicity $%
t_{1}\leqslant t\leqslant t_{2}$ $(z_{t}\geqslant 0)$, $t_{2}\leqslant
t\leqslant t_{3}$ $(z_{t}\leqslant 0)$, etc. According to (\ref{Eq36}) and (%
\ref{eq38}) we can write%
\begin{eqnarray}
\underset{t_{1}\leqslant t\leqslant t_{2}}{X_{\pm }^{2}(t)} &=&X_{\pm
}^{2}(t_{1})\exp \int_{\zeta _{1}}^{z}\frac{f_{\mp }}{w\sqrt{g}}\, dz, 
\notag \\
\underset{t_{2}\leqslant t\leqslant t_{3}}{X_{\pm }^{2}(t)} &=&X_{\pm
}^{2}(t_{2})\exp \int_{\zeta _{2}}^{z}\frac{f_{\pm }}{w\sqrt{g}}\, dz,%
\mathrm{\ etc.}  \label{eq43}
\end{eqnarray}%
To find the characteristic exponent $\mu $ consider, e.g., the growing
solution $X_{+}(t)$. Setting $t=t_{2}$, $z(t_{2})=\zeta _{2}$ in the first
equation of (\ref{eq43}) and $t=t_{3}$, $z(t_{3})=\zeta _{1}$ in the second
one, we can express $X_{+}^{2}(t_{3})$ through $X_{+}^{2}(t_{1})$. Using Eq.
(\ref{eq22}) and taking into account that $t_{3}=t_{1}+\tau $, $\varphi
(t+\tau )=\pm \varphi (t)$, we thus obtain%
\begin{equation}
\mu =-\frac{W}{\tau }\int_{\zeta _{1}}^{\zeta _{2}}\frac{dz}{w\sqrt{g}}.
\label{eq44}
\end{equation}%
Recall that $\tau $ is the period of $z(t)$, the constant $W$ is determined
from Eq. (\ref{eq41}) or Eq. (\ref{eq42}), its sign being taken opposite to
that of the integral in (\ref{eq44}) to provide for positivity of $\mu $.
Since $w(z)$ has zeros, the integrals in Eqs. (\ref{eq43}) and (\ref{eq44})
are understood as their principal values. Formula (\ref{eq44}) is a simple
generalization of the ones used previously in Refs. \cite%
{Greene,Kaiser,Maslov4,Finkel}.

\subsection{Evaluation of the characteristic exponent\textit{\ }}

Let us return to Eq. (\ref{eq19}). It can be written in the form of Eq. (\ref%
{eq21}) if we set%
\begin{eqnarray}
z(t) &=&-\ln (a/a_{\max })^{2},\quad z(0)=0,  \label{eq45} \\
h(z) &=&E-3+\omega _{0}^{2}+z.  \label{eq48}
\end{eqnarray}%
Equation (\ref{eq7}) then immediately gives%
\begin{equation}
g(z)=4[\omega _{0}^{2}(e^{z}-1)-z].  \label{eq47}
\end{equation}%
Zeros of this function are shown in Fig. 3. Since $\xi (t)=a/a_{\max }$
oscillates with the period $T$ in the interval $-1\leqslant \xi \leqslant 1$
[see Eqs. (\ref{eq7})-(\ref{eq12})], the function $z(t)$ (\ref{eq45})
oscillates with the period $\tau =T/2$ between the turning points $\zeta
_{1}=0$ and $\zeta _{2}=+\infty $. 
\begin{figure}[th]
\centering\includegraphics[width=6cm]{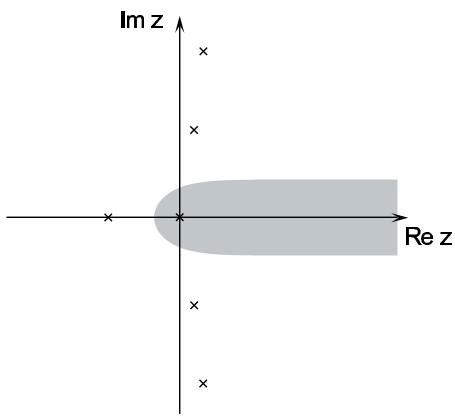}
\caption{The layout of zeros of $g(z)$ (\ref{eq47}) on the complex $z$
plane. The vicinity $D$ (shaded) of the interval $[0,\infty )$ belongs to
the domain of analyticity of $w(z)$.}
\end{figure}
To calculate the characteristic exponent by the formula (\ref{eq44}) we need
to know the function $w(z)$ which is the solution of Eq. (\ref{eq27}) with
boundary conditions (\ref{eq26}). Unfortunately, for given $h(z)\ $(\ref%
{eq48}) and $g(z)$ (\ref{eq47}) equation (\ref{eq27}) cannot be solved
analytically. We solve it numerically for various values of the parameters $E
$ and $\omega _{0}^{2}=1-\ln a_{\max }^{2}$ \cite{Koutvitsky-Maslov}. Doing
so, we use the conditions (\ref{eq26}) in the following way. As discussed
above, the fulfilment of (\ref{eq26}) at a finite turning point means
analyticity of $w(z)$ in some vicinity of this point. Therefore, we can use
the expansions (\ref{eq28a}) setting there $\zeta =\zeta _{1}=0$, $g_{0}=0$, 
$g_{1}=4(\omega _{0}^{2}-1)$, $g_{n}=4\omega _{0}^{2}/n!$ $(n=2,3,\dots )$, $%
h_{0}=E+\omega _{0}^{2}-3$, $h_{1}=1$. Equation (\ref{eq33}) then gives%
\begin{equation}
w_{2}=-\frac{1+(2E+3\omega _{0}^{2}-6)w_{1}}{6(\omega _{0}^{2}-1)}.
\label{eq49}
\end{equation}%
We thus solve Eq. (\ref{eq27}) with the following conditions at $z=0$: $%
w(0)=1$, $w^{\prime }(0)=w_{1}$, $w^{\prime \prime }(0)=2w_{2}$. Given
values of $E$ and $\omega _{0}^{2}$, we choose $w_{1}$ so as to satisfy the
condition (\ref{eq26}) at infinity,%
\begin{equation}
\left( w^{\prime }\sqrt{g}\right) _{z\rightarrow +\infty }\rightarrow 0.
\label{eq50}
\end{equation}%
At the same time, since $\mu $ assumed to be real, the values of $E$, $%
\omega _{0}^{2}$, and $w_{1}$ must provide for positivity of $W^{2}$,%
\begin{equation}
W^{2}=4[3-E-\omega _{0}^{2}-(\omega _{0}^{2}-1)w_{1}]>0  \label{eq51}
\end{equation}%
[see Eqs. (\ref{eq41}) and (\ref{eq44})]. Conditions (\ref{eq50}) and (\ref%
{eq51}) determine the resonance zones in the space of parameters $E$ and $%
\omega _{0}^{2}$ (or $a_{\max }^{2}$). Hereinafter the zones will be
referred to as $Z_{j}$ and numbered sequentially as $E$ grows (with $a_{\max
}^{2}$ fixed) starting with $j=-1$ in the region $E<0$. Figure 4 shows the
solutions $w(z)$ for zones $Z_{1}$, $Z_{2}$, and $Z_{3}$ lying in the region 
$E>2$. 
\begin{figure}[th]
\includegraphics[width=8cm]{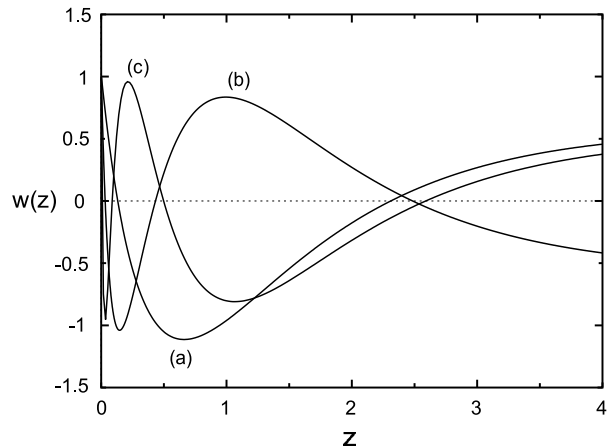}
\caption{Behavior of $w(z)$ for $E=4$ in the different resonance zones: (a)
zone $Z_{1}$, $a_{\max }^{2}=0.5625$, (b) zone $Z_{2}$, $a_{\max }^{2}=0.9025
$, (c) zone $Z_{3}$, $a_{\max }^{2}=0.9598$. The values of $a_{\max }^{2}$
choosen correspond to the centers of zones where $\protect\mu $ achieves its
maxima. One can see that the number of zeros of $w(z)$ is unit above the
number of a zone.}
\end{figure}
Now, knowing $w(z)$, we can calculate the integral in (\ref{eq44}). Because $%
w(z)$ has zeros, we first transform the integrand with the help of Eq. (\ref%
{eq40}) extracting the total derivative. Owing to the condition (\ref{eq50}%
), the latter does not contribute to the principal value of the integral,
while the remaining terms give 
\begin{equation}
\int_{0}^{\infty }\frac{dz}{w\sqrt{g}}=-\frac{1}{2W^{2}}\int_{0}^{\infty }%
\left[ \sqrt{g}\left( w^{\prime }\sqrt{g}\right) ^{\prime }\ln w^{2}+8hw%
\right] \frac{dz}{\sqrt{g}}.  \label{eq52}
\end{equation}%
\begin{figure}[th]
\includegraphics[width=8cm]{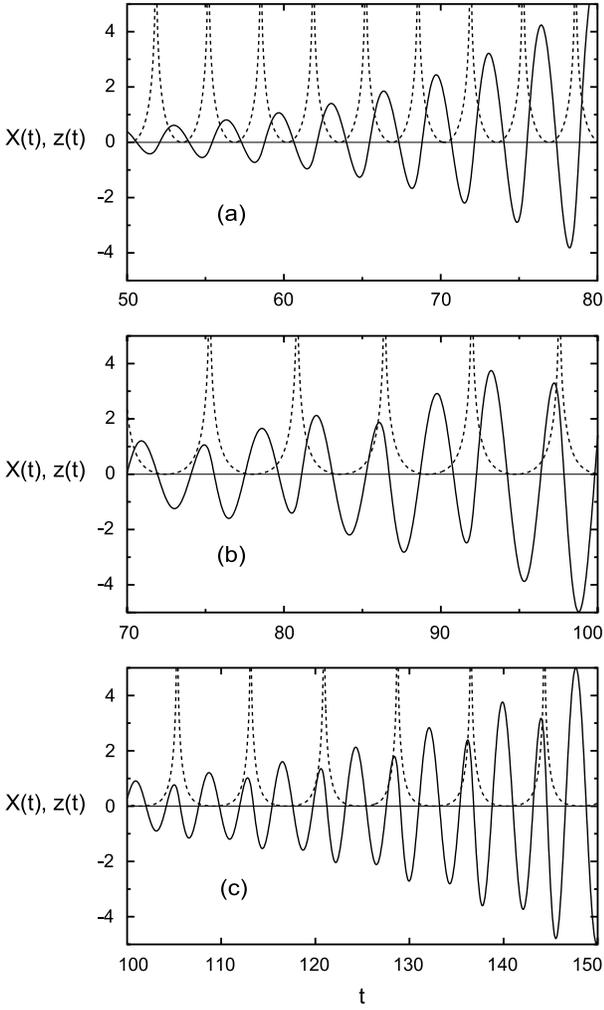}
\caption{Resonant solutions of Eq. (\ref{eq19}), $X(t)\sim X_{+}(t)=\protect%
\varphi (t)e^{\protect\mu t}$, (solid lines) and the function $z(t)=-\ln
(a/a_{\max })^{2}$ (dashed lines): (a) zone $Z_{1}$, (b) zone $Z_{2}$, (c)
zone $Z_{3}$. The initial conditions are: $a(0)=a_{\max }$, $a_{t}(0)=0$, $%
X_{t}(0)=0$, $X(t)$ is normalized in a proper way. The values of $E$ and $%
a_{\max }^{2}$ in each zone are the same as in Fig 4. It is seen that $%
\protect\varphi (t)$ is $\protect\tau $-periodic in $Z_{1}$, $\protect\tau $%
-antiperiodic in $Z_{2}$, $\protect\tau $-periodic in $Z_{3}$, and so on, in
accordance with the solutions (\ref{eq43}) [see Eqs. (\ref{eq39}) and (\ref%
{eq42}) and Fig. 4].}
\end{figure}
The integrand on the right-hand side of Eq. (\ref{eq52}) is more convinient
for numerical integration because its singularities are all integrable. We
perform the integration in (\ref{eq52}), calculate $W^{2}$ by the formula (%
\ref{eq51}), and find the period $T=2\tau $ by integration of Eq. (\ref{eq7}%
). These procedures are carried out numerically for a set of grid points in
every resonance zone. In this way from (\ref{eq44}) we obtain the
characteristic exponent $\mu $ as a function of $E$ and $a_{\max }^{2}$.

To check this result we derive $\mu (E,a_{\max }^{2})$ directly from
analysis of numerical solutions of Eq. (\ref{eq19}). Examples of these
solutions for resonance zones $Z_{1}$, $Z_{2}$, and $Z_{3}$ are shown in
Fig. 5. The growth of the amplitude with time is clearly seen. The function $%
\mu (E,a_{\max }^{2})$ so derived is found to be fully coincident with the
one obtained by the formula (\ref{eq44}). 
\begin{figure}[ht]
\includegraphics[width=8.5cm]{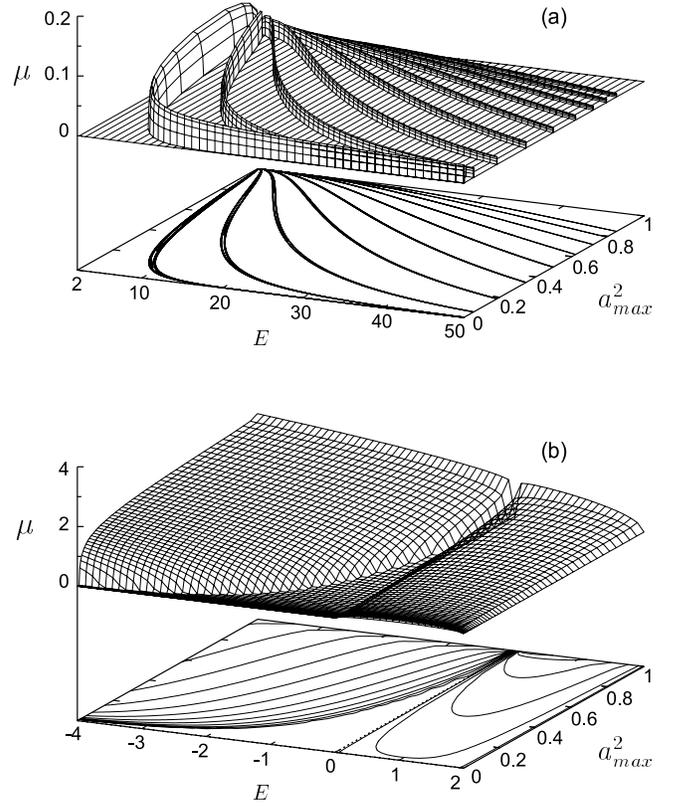}
\caption{The stability-instability chart: (a) $E>2$, first ten zones are
shown, (b) $E\leqslant 2$.}
\end{figure}

The resulting stability-instability chart is presented in Fig. 6. Figure
6(a) shows the region $E>2$. There is an infinite series of narrow resonance
zones $Z_{1}$, $Z_{2}$, $Z_{3}$, \dots , the first one having the highest
magnitude ($\approx 0.08$ at the maximum). All these zones originate from
the point $E=2$, $a_{\max }^{2}=1$ at which $\mu =0$ [see Eqs. (\ref{eq44})
and (\ref{eq51})]. In the region $E\leqslant 2$ we have two zones, $Z_{0}$
and $Z_{-1}$, lying in the ranges $0<E<2$ and $E<0$, correspondingly. Since
in these zones the values of $\mu $ proved to be much greater than in $Z_{1}$%
, $Z_{2}$, \dots , we depict the surface $\mu (E,a_{\max }^{2})$ for this
region separately, in Fig. 6(b).

\subsection{Spatial structure of the perturbation}

Consider now Eq. (\ref{eq20}). It has the form of the Schr\"{o}dinger
equation for a quantum particle of the energy $E$ moving in the\ potential $%
-\ln u^{2}$. Since the potential tends to $+\infty $ with growing $r$ [as $%
r^{2}$, see Eqs. (\ref{eq13}) and (\ref{eq13a})], the energy spectrum is
discrete, $E=E_{n}$, and the corresponding eigenfunctions $\Psi _{n}(\mathbf{%
r})$ are all localized. In the case of the nodeless pulson (\ref{eq13}) we
have the isotropic harmonic oscillator. Its eigenfunctions are well known
(see, e.g., \cite{Flugge}). We write them as follows:%
\begin{eqnarray}
\Psi _{n}(\mathbf{r}) &=&\sum_{l=0}^{n}[1+(-1)^{n-l}]R_{nl}(r)Y_{l}(\theta
,\varphi ),  \label{eq53} \\
R_{nl}(r) &=&r^{l}e^{-r^{2}/2}\Phi \left( -\frac{n-l}{2},\ l+\frac{3}{2},\
r^{2}\right) ,  \label{eq54} \\
Y_{l}(\theta ,\varphi ) &=&\sum_{m=-l}^{l}c_{l,m}P_{l}^{\left| m\right|
}(\cos \theta )e^{im\varphi }.  \label{eq55}
\end{eqnarray}%
Here $\Phi (\alpha ,\gamma ,x)$ is the Kummer function, $P_{\nu }^{\mu }(x)$
are the associated Legendre functions, $c_{l,m}$ are constants, $%
c_{l,-m}=c_{l,m}^{\ast }$. The energy spectrum is given by%
\begin{equation}
E=E_{n}=2n\qquad (n=0,1,2,\dots).  \label{eq56}
\end{equation}%
(Our energy levels are shifted with respect to the conventional ones since
the minimum of the potential $-\ln u_{0}^{2}$ is $-3$.)

In the case of the nodal pulsons the picture becomes more complicated due to
the loss of the orbital degeneracy. The corresponding eigenfunctions and
eigenvalues can be calculated only numerically. As an example, in Fig. 7 is
shown the energy spectrum for perturbations of the one-nodal pulson. 
\begin{figure}[th]
\centering\includegraphics[width=8 cm]{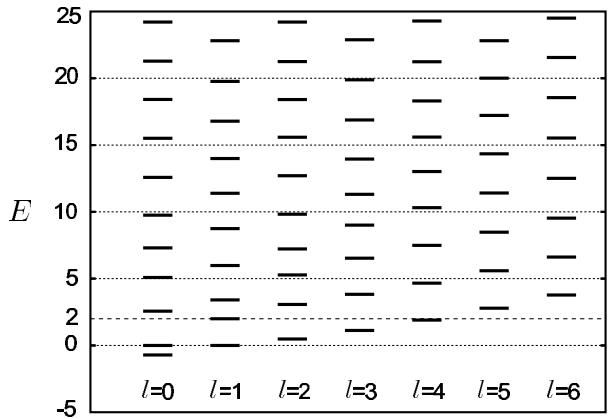}
\caption{The energy levels for the perturbations with different orbital
numbers $l$. The case of the one-nodal pulson.}
\end{figure}

Note that there always exist the eigenvalues $E=0$, $l=0$ and the
corresponding eigenfunction $\Psi _{0}(r)\varpropto u(r)$. This fact
immediately follows from the comparison of Eqs. (\ref{eq6}) and (\ref{eq20}%
). The corresponding $X_{0}(t)$ in $\eta (t,\mathbf{r})$ is an oscillating
function with the amplitude growing linearly with time. It is easy to see,
however, that this mode is physically meaningless. Indeed, it will formally
appear if we perturb the pulson by a small variation of its amplitude $%
a_{\max }$ but not the form $u(r)$. Due to nonlinearity, this results in a
pulson with slightly shifted frequency. Then the difference of the perturbed
and unperturbed pulsons, i.e., $\eta (t,r)$, will have the form of beats
generated by two oscillations with close frequencies and the same profile $%
u(r)$. The function $X_{0}(t)$ approximates the initial, linearly growing
part of a beat. We exclude this mode from the subsequent consideration,
since it belongs to the class of perturbations that conserve a pulson as a
whole. Next, for the nodal pulsons only, there is a mode with $E=0$, $l=1$
(see Fig. 7). Since this mode cannot grow faster than linearly in time, we
also do not take it into account. Further, we should exclude the mode
resulting from a small translation of the pulson. The corresponding
eigenfunction is proportional to $\mathbf{n}\nabla u$, where $\mathbf{n}$ is
a displacement vector. Using Eqs. (\ref{eq6}) and (\ref{eq20}) one can
easily show that this mode corresponds to $E=2$, $l=1$. Thus the resulting
perturbation is written as%
\begin{equation}
\eta (t,\mathbf{r})=\sum\limits_{n}X_{n}(t)\Psi _{n}(\mathbf{r}),
\label{eq31}
\end{equation}%
where $X_{n}$ is a solution of Eq. (\ref{eq19}) with $E=E_{n}$, $E_{n}\neq
0,2$. If $E_{n}$ and $a_{\max }^{2}$ are in a resonance zone, $X_{n}(t)$
will be represented as a linear combination of the solutions (\ref{eq22})
and, hence, will grow with time as $e^{\mu (E_{n},a_{\max }^{2})t}$.

\subsection{Instability of the pulsons}

The arrangement of the resonance zones on the $(E,a_{\max }^{2})$ plane
indicates that for any spectrum $E_{n}$ there always exist the ranges of $%
a_{\max }^{2}$ where pulsons are unstable. But do the values of $a_{\max
}^{2}$ exist for which the pulsons are stable? To answer this question let
us return to the surface $\mu (E,a_{\max }^{2})$ depicted in Fig. 6. Take,
at first, the spectrum for the nodeless pulson. We choose the sections $\mu
_{n}(a_{\max }^{2})$ of the surface $\mu (E,a_{\max }^{2})$ by $E=2n$ and
project them on the $(\mu ,a_{\max }^{2})$ plane. As a result, the pattern
shown in Fig. 8(a) emerges. 
\begin{figure}[th]
\centering\includegraphics[width=8cm]{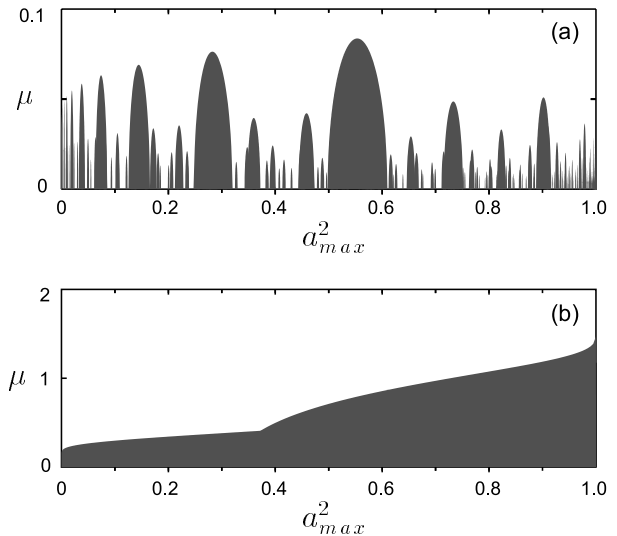}
\caption{Superposition of the sections $\protect\mu _{n}(a_{\max }^{2})$ of
the surface $\protect\mu (E,a_{\max }^{2})$: (a) the nodeless pulson, $%
E_{n}=4$, $6$, $8$, \dots , $50$ $(n=2,3,4,\dots ,25)$, (b) the one-nodal
pulson, $E_{n}=-0.7142$, $0.4833$, $1.1222$, $1.8996$, \dots .}
\end{figure}
It is clearly seen the tendency to the total filling of the interval $%
0<a_{\max }^{2}<1$ by the resonant peaks as the successively higher energy
levels are accounted for. This implies that for any given $a_{\max }^{2}$
there always exists an unstable mode with $\mu =\mu _{n}(a_{\max }^{2})$,
i.e., strictly speaking, all nodeless pulsons are unstable. On the other
hand, the figure shows that there are domains of $a_{\max }^{2}$ where the
peaks are very small. These domains are the gaps between the main peaks
originated from the low-energy cross sections of the surface $\mu (E,a_{\max
}^{2})$ over a few first zones. In the gaps the exponent $\mu $ is small, so
that the corresponding pulsons are long-lived. For example, in Ref. \cite%
{Koutvitsky-Maslov} we observed numerically the nodeless pulson with $%
a_{\max }^{2}=0.49$ that conserved its coherency against the radially
symmetric perturbations over the course of several hundreds of periods.

Further, the above projective procedure is performed using the spectrum of
the one-nodal pulson (Fig. 7). The main contribution here is made by the
sections with the energies $E_{n}=-0.7142$ $(l=0)$, $0.4833$ $(l=2)$, $1.1222
$ $(l=3)$, and $1.8996$ $(l=4)$ falling into zones $Z_{-1}$ and $Z_{0}$, the
projections of the first and the third sections overlapping the other ones.
The result is presented in Fig. 8(b). We see that $a_{\max }^{2}$ axis is
totally full. Thus, the one-nodal pulson has neither stability nor even
quasistability domains. It seems likely that things will get worse, not
better, if one goes to the multinodal pulsons. We thus conclude that,
strictly speaking, all pulsons of the model considered are unstable. But
nodeless pulsons can be quasistable in narrow ranges of amplitudes. It is
the long-lived pulsons that can be of astrophysical and cosmological
interest. If the dark matter consists of scalar particles, such pulsons will
be realistic candidates for the dark matter objects having oscillating
density \cite{Seidel}.

\subsection{On the instability of the Affleck-Dine type condensate}

The obtained stability-instability chart turns out to be appropriate for the
stability analysis of the nonlocalized coherent states as well. As an
example, we consider a uniformly distributed background $\phi _{0}(t)$, a
scalar condensate, oscillating around the minimum of the potential (\ref{eq4}%
) at $\phi =0$. This state can be formally obtained from Eq. (\ref{eq2}) if
we set there $u(\mathbf{r})\equiv 1$. We thus assume that $\phi _{0}(t)$
obeys Eq. (\ref{eq5}). Taking the perturbed state $\phi =\phi _{0}(t)+\eta
(t,\mathbf{r})$, in the linear approximation from Eqs. (\ref{eq1}) and (\ref%
{eq4}) we readily obtain%
\begin{equation}
A_{tt}+(k^{2}-2-\ln \phi _{0}^{2})A=0,  \label{eq63}
\end{equation}%
where $A(t,\mathbf{k})$ is the Fourier amplitude of the perturbation, and $%
k=\left| \mathbf{k}\right| $. It is seen that the real and imaginary parts
of this equation have the form of Eq. (\ref{eq21}) with $h(z)$ given by Eq. (%
\ref{eq48}), $z=-\ln (\phi _{0}/\phi _{0\max })^{2}$, $\omega _{0}^{2}=1-\ln
\phi _{0\max }^{2}$, and $E=k^{2}$. Returning to the stability-instability
chart (Fig. 6) we note that in the region $E\geqslant 0$ maximal values of $%
\mu $ are attained in the zone $Z_{0}$ for which $0<E<2$. Interestingly,
this band exactly coincides with the one obtained in Ref. \cite{Enqvist3}
for the power-law potential approximating (\ref{eq3}) when $\lambda \ll 1$.
In the interior of $Z_{0}$ the exponent $\mu $ depends almost not at all on
the amplitude of the condensate oscillations and is a sufficiently smooth
function of $k^{2}$ with a maximum at $k^{2}=k_{0}^{2}\approx 1$ where $\mu
\approx 0.5$. Therefore, if the initial power spectrum $\left| A(0,\mathbf{k}%
)\right| ^{2}$ lies in the region $0\lesssim \left| \mathbf{k}\right|
\lesssim \sqrt{2}$ and, in addition, its characteristic width along $\mathbf{%
k}$ is small, $\Delta k\ll \sqrt{2}$, then the growth of the perturbation
amplitude will not be accompanied by significant changes in the structure of
the perturbation. The limiting case of such perturbations is a harmonic
wave. Otherwise, if $\Delta k\gtrsim \sqrt{2}$, the shape of the power
spectrum will vary with time so that a maximum will appear at $k_{0}\approx 1
$. As a result, the effective width of the spectrum will become smaller, $%
\Delta k\lesssim 1$. In this case, if the initial spectrum is sufficiently
isotropic in $\mathbf{k}$ space, the parametric amplification of the
perturbations will result in the emergence of the localized field
configurations of the characteristic size $\Delta r\sim 1/\Delta k\gtrsim 1$
that agrees with the radius of the gausson (see Sec. II). At this scale the
field practically does not undergo spatial oscillations since the
corresponding wavelength $2\pi /k_{0}\gtrsim 1$. We thus expect that at the
nonlinear stage these configurations will turn into the nodeless pulsons.
Their period will be equal to the period of the condensate oscillations
since in the zone $Z_{0}$ the parametric amplification proceeds at the basic
frequency. Gradually, the energy of the oscillating condensate will go to
ensemble of the arising pulsons, this process resulting in the damping of
the background oscillations. As to the pulsons themselves, they can be
long-lived or short-lived depending on their amplitudes, in accordance with
the results of the previous Subsection.

Note, that numerical simulations performed for the complex version of the
model (\ref{eq1}) and (\ref{eq3}) have shown the fragmentation of both the
rotating \cite{Kasuya, Enqvist2} and oscillating \cite{Enqvist3}
Affleck-Dine condensate. The localized configurations arising in the
condensate have been identified with Q-balls. We belive, however, that the
configurations observed in the oscillating condensate are in fact the
complex pulsons (see Sec. IV), rather than the usual Q-balls. This
possibility was early discussed in Ref. \cite{Enqvist4} where an attempt to
simulate the complex pulson has been made.

The resonant exitation of the pulsons was also observed in the two-vacuum $%
\phi ^{4}-\phi ^{6}$ model within a regularly oscillating background \cite%
{Maslov4} and in the $\phi ^{4}$ model within an initially thermalized
background \cite{Gleiser2}. Note that in two-vacuum models the pulsons can
play the role of nuclei of a new phase. In Ref. \cite{Maslov4} the general
suggestion has been made that the parametric resonance can underlie the
mechanism responsible for the first-order phase transitions in nonlinear
non-dissipative systems. This conjecture turns out to be in agreement with
recent results of Ref. \cite{Gleiser3} where the resonant nucleation within
the thermalized background have been numerically observed in the $\phi
^{3}-\phi ^{4}$ model. Note, in addition, that the dynamical nucleation can
also take place in the nonlinear Schr\"{o}dinger equation \cite{Barashenkov}.

\section{CONCLUDING REMARKS}
In this paper we have examined only the linear stage of instability at which
small deformations of the pulson's shape result in loss of the coherence.
There is numerical evidence that in time the growth of the perturbations
becomes saturated due to nonlinear effects \cite{Koutvitsky-Maslov}. We thus
suggest that in the model considered the pulsons, while unstable, remain
well localized objects with no tendency for spreading or collapsing.

Further, we dealt with a real scalar field. It would be interesting to
perform the similar analysis for a complex scalar field too. It is believed
that the existence of the scalar charge can stabilize a field lump. For
Q-balls this fact is well established (so-called Q-theorem \cite{Lee,
Rybakov, Belova}). In contrast, for the complex pulsons this is an open
question. As it was shown in Refs. \cite{Marques,Bogolubsky}, the field
equation (\ref{eq1}) with $U^{\prime }=-\phi \ln (\phi \phi ^{\ast })$
admits the exact pulson solutions of the form $\phi
_{0}(t,r)=a(t)u(r)e^{i\theta (t)}$, where $a(t)$, $u(r)$, and $\theta (t)$
are real. The function $u(r)$ satisfies Eq. (\ref{eq6}) as before, while $%
a(t)$ oscillates with a period $T$ in accordance with the equation 
\begin{equation}
a_{tt}=-\frac{d}{da}\left[ \frac{1}{2}a^{2}(1-\ln a^{2})+\frac{q^{2}}{2a^{2}}%
\right] ,  \label{eq64}
\end{equation}%
where $q$ is a real constant, $q^{2}<(2e)^{-1}$, and $\theta _{t}=qa^{-2}$.
The constant $q$ is proportional to the charge of the scalar field. In
contrast to Eq. (\ref{eq5}), the potential in the square brackets of Eq. (%
\ref{eq64}) prevents $a(t)$ from being zero. Without loss of generality one
may assume $a(t)$ positive, so that the oscillations occur arround the
minimum of the potential at $a=a_{0}$, where $a_{0}$ is the least positive
root of the equation $a^{4}\ln a^{2}=-q^{2}$. If $a$ is at rest in this
minimum, then $\theta (t)=qa_{0}^{-2}t+\theta (0)$, and we have the standard
Q-ball. Physically, Eq. (\ref{eq64}) describes the motion of a mechanical
particle with an angular momentum $q$ in the potential $(a^{2}/2)(1-\ln
a^{2})$ \cite{Landau}. The condition for its trajectory to be closed is $%
\theta (T)-\theta (0)=2\pi m/n$, where $m$ and $n$ are arbitrary integers.
In fact, it relates the energy of the particle and its angular momentum
whereby such trajectories exist. In our case this means periodicity of the
solution $\phi _{0}(t,r)$ with the period $nT$. Obviously, there is an
infinity of such solutions. Taking $\phi _{0}(t,r)$ and considering the
partial perturbation $\eta \varpropto X(t)\Psi (\mathbf{r})$ one can find
that the function $\Psi (\mathbf{r})$, assumed to be real, satisfies Eq. (%
\ref{eq20}) as before, while $X(t)$ obeys the equation 
\begin{equation}
X_{tt}+(E-1-\ln a^{2})X=e^{2i\theta }X^{\ast },  \label{eq65}
\end{equation}%
where $E$ is a real constant. This equation can be represented as a system
of four real first-order equations with periodic coefficients of the periods 
$T$ and $nT$. It is significant that, since $a(t)\neq 0$, these coefficients
are bounded in time, so that one can attempt to estimate the characteristic
exponent of the system using the standard methods \cite{Yakubovich}.

Also, it would be interesting to examine stability of a selfgravitating
pulson. Hopefully, gravitation can expand the domains of (quasi)stability,
as it is the case for Q-balls \cite{Rybakov}.These are possible subjects of
our future work.

In the present paper we have investigated stability of both the coherent
localized states (pulsons) and nonlocalized states (uniformly oscillating
scalar condensate) of the real scalar field. Our main analytical result is
the generalization of the Lindemann-Stieltjes method to the case that the
periodic coefficient in the Hill equation is unbounded in time. Our main
numerical result is the stability-instability chart with the values of
characteristic exponent calculated in the resonance zones. Using this chart
we have found the gaps in the set of the pulson amplitude values in which
the real nodeless pulsons conserve the coherency for an extremely long time.
Also, considering the oscillating scalar condensate, we have detemined the
wavelength of the most unstable mode. This wavelength turned out to be equal
to the characteristic size of the nodeless pulson. We thus suggest the
pulsons can be formed due to resonant fragmentation of the scalar
condensate. These are our main physical results.

\acknowledgments {\ The authors thank I. Bogolubsky, Yu.P. Rybakov, and A.
Shagalov for useful discussions. This work was partly supported by the RAS\
Presidium Program ``Nonstationary phenomena in astronomy''. }

\end{document}